\documentclass[conference]{IEEEtran}
\pdfoutput=1
\usepackage{dsfont}
\usepackage{graphicx}
\usepackage{amsmath}
\usepackage{amsfonts}
\usepackage{amssymb}

\ifCLASSINFOpdf

\else

\fi

\begin{document}

%\title{Scheduling in Cognitive Radio Networks\\ Using Belief Propagation}
\title{Message-Passing Algorithms for Optimal Utilization \\of Cognitive Radio Networks}

\author{\IEEEauthorblockN{Hamed Mahmoudi}
\IEEEauthorblockA{
%The Nonlinearity and Complexity
%\\Research Group, Aston University,\\ Birmingham B4 7ET, United Kingdom
%\\
Aston University, UK\\
 h.mahmoudi@aston.ac.uk}
\and
\IEEEauthorblockN{Georgios Rodolakis}
\IEEEauthorblockA{
%Institute of Telematics and Informatics\\
%Volos, Greece\\
CERTH, Greece\\
rodolakis@iti.gr}
\and

\IEEEauthorblockN{Leonidas Georgiadis}
\IEEEauthorblockA{
Aristotle University of Th., Greece\\
leonid@auth.gr}
\and
\IEEEauthorblockN{David Saad}
\IEEEauthorblockA{
%The Nonlinearity and Complexity \\Research Group, Aston University,\\ Birmingham B4 7ET, United Kingdom
%\\
Aston University, UK\\
d.saad@aston.ac.uk}

}

\maketitle

\begin{abstract}
%\boldmath
Cognitive Radio has been proposed as a key technology to significantly improve spectrum usage in wireless networks by enabling unlicensed users to access unused resource.
We present new algorithms that are needed for the implementation of opportunistic
scheduling policies that maximize the throughput utilization of resources by secondary users, under maximum interference constraints imposed by existing primary users.
Our approach is based on the Belief Propagation (BP) algorithm, which is advantageous due to its simplicity and potential for distributed implementation.
We examine convergence properties and evaluate the performance of the proposed BP algorithms via simulations and  demonstrate that the results compare favorably with a benchmark greedy strategy.
\end{abstract}

\IEEEpeerreviewmaketitle

\section{Introduction}
Cognitive Radio Networks (CRNs) have recently been proposed as a promising approach to improve spectrum usage, by enabling unlicensed users to access unused wireless resources~\cite{akyildiz2006next}.
The key technology is cognitive radio
%~\cite{mitola2000software}
that can
dynamically adjust its channel allocation depending on spectrum availability.
Therefore, in CRNs, we distinguish between: Primary Users (PU), who
have licensed rights to network resources,  and (unlicensed) Secondary Users (SU) equipped with cognitive radios, who access underutilized spectrum channels.
A basic requirement is that secondary users must not adversely impact the performance of  primary users.
This can be ensured by a coordination mechanism, under the
terms of an agreement between users, but in many practical cases such extensive coordination is
unfeasible.
In contrast, an approach that leverages the cognitive radio capabilities consists in secondary
users monitoring the spectrum and exploiting opportunistically 
channels unemployed by primary users.
%free primary channels.
% (i.e., licensed communication channels that are unused by their primary users for a certain time).
Combined with the dynamic wireless environment this poses important challenges as the channel access algorithms must be very efficient for the access to be updated continuously, and distributed to allow for uncoordinated opportunistic operation.

In this paper, we present new algorithms needed for opportunistic
scheduling, which maximize the utility
of the secondary users,  under maximum interference constraints imposed by existing primary users.
The utility can be a function of actual or stochastic traffic and communication rates, or corresponding queue lengths; such choices can be formulated with the goal to maximize the achievable network throughput using Lyapunov Optimization techniques (see the seminal paper~\cite{tassiulas1992stability},
% in the context of multi-hop networks,
and~\cite{urgaonkar2008opportunistic} for an application in CRNs).
Our aim is to select sets of links between secondary users and free primary channels that can be activated simultaneously, such as to maximize the total utility of secondary users, without causing disruptive interference to existing primary users.
Hence, the optimization problem in our framework, which we describe in Section~\ref{sec:model}, is a generalization of Maximum Weight Matching (MWM), with additional interference constraints. 

Our approach is based on the Belief Propagation (BP) algorithm, which is advantageous due to its simplicity and potential for distributed implementation.
BP is an iterative message-passing algorithm, that was discovered independently in the information theory~\cite{GallagerMonograph}, machine learning~\cite{pearl1988probabilistic} and statistical physics~\cite{MPV} communities.
%%%%%%%%%%%%%%%%%%%%%%%%%
It has been used, with spectacular experimental success, in many application areas, for instance iterative decoding
and combinatorial optimization, which involve graphs with cycles.
The physical
interpretation of message-passing algorithms and in particular belief propagation was
later explained by statistical mechanics studies of disordered systems~\cite{KStap,yedidia2001generalized,mezard2009information}, consolidating BP and its generalizations to a very powerful tool that provides practical solutions to hard problems. Additionally, BP has been shown to solve the classical Maximum Weight Matching exactly in bipartite graphs with cycles~\cite{bayati2008max}.

Cognitive radio networks have attracted significant research interest (see~\cite{akyildiz2006next} for a survey), including work on optimal spectrum
scheduling. However, the problem of efficient and fully distributed throughput optimization under interference remains challenging.
A recent paper~\cite{shamaiah2012distributed} proposes affinity propagation based algorithms for spectrum
access in CRNs, in the specific case where primary users permit secondary users access as long as they consent to act as relays.
 The study that is closest to our approach is~\cite{urgaonkar2008opportunistic}, which introduces a general framework using the Lyapunov Optimization technique to
design a scheduling policy for flow control and resource allocation in CRNs.  It motivates the formulation of the optimization problem we aim to address; specifically, the policy proposed in~\cite{urgaonkar2008opportunistic} is required to solve a certain deterministic optimization problem at each time slot, which imposes a computational bottleneck when interference is taken into account. In this paper, our contribution is complementary as we focus on a more general algorithmic solution, and the problem we address can 
be used to solve the optimization problem discussed above,  in contrast to~\cite{urgaonkar2008opportunistic} that merely presents a greedy algorithm for the simplified case of no interference.

The paper is organized as follows: we present a general modeling framework for opportunistic cognitive radio networks and formulate the problem of optimal channel allocation with interference control in Section~\ref{sec:model}; in Section~\ref{sec:BP}, we develop two new Belief Propagation algorithms, which can be fully decentralized, to solve the optimization problem for hard and soft interference constraints, respectively; in Section~\ref{sec:simulations}, we show the convergence and evaluate the performance of the proposed BP algorithms via simulation scenarios in a realistic Signal to Interference plus Noise (SINR) based model of cognitive radio networks and demonstrate that the results compare favorably with a benchmark greedy scheduling strategy.

%Although primary users prohibit secondary users to access channels, a mechanism can be designed to enable sharing of the spectrum under the
%terms of an agreement between primaries and secondaries.
%
%
%
%%\section{Cognitive Radio Networks}
%
%Cognitive radios (CRs) are a promising technology for enabling
%unlicensed devices to efficiently use the extra channel capacities.
%They have been proposed to
%improve spectrum efficiency by having the cognitive radios act
%as secondary users (compared to primary users) to opportunistically access extra frequency bands.
%
%Therefore in a cognitive radio networks (CRNs) we can distinguish two types of users : primary, those which
%have priority rights to access channel frequencies and secondary (or cognitive) that are users which can only use channels if they are not occupied by primaries.
%Hence secondary users need to constantly sense the presence of the primary users and to detect possible spectrum holes.
%
%Although primary users prohibit secondary users to access channels, a mechanism can be designed to enable sharing of the spectrum under the
%terms of an agreement between primaries and secondaries.

\section{Model}
\label{sec:model}
\begin{figure}[htb]
\centering
\hspace{0.4mm}
\includegraphics[width=0.55\columnwidth]{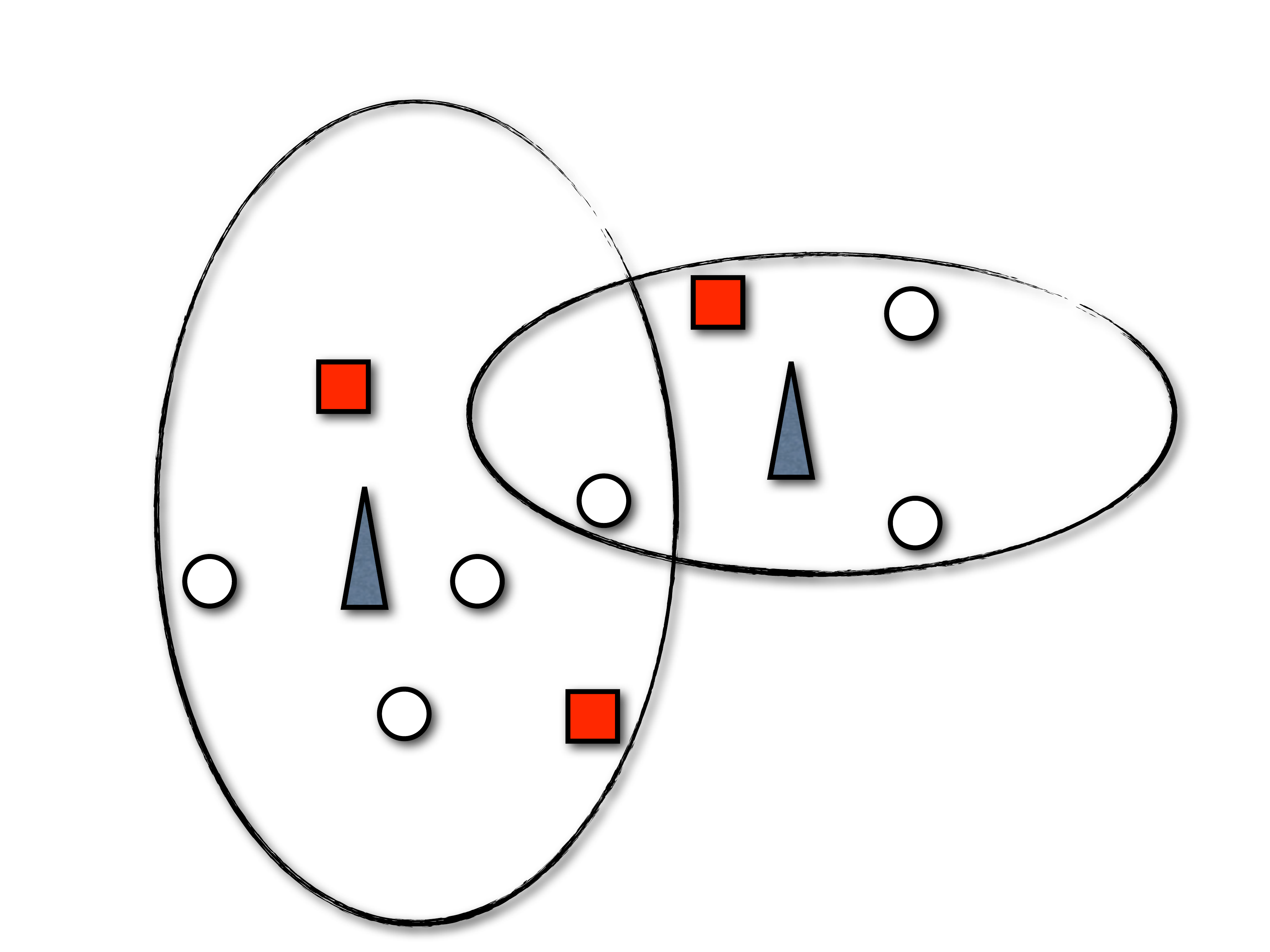}
\caption{ A simple cognitive radio network, with two channels to serve users.
Primary users are shown in red, and cognitive users are
while circles. }
\label{fig:cognitive1}
\end{figure}

The model adopts a hierarchical access structure with primary and secondary users. The basic idea is to
open channel frequencies to secondary users while considering the interference experienced by primary users
(licensees). Fig~\ref{fig:cognitive1} illustrates a CRN. Primary users (red colored) are first to be served by channels.
Secondary users, in non-cooperative scenario can not access channels if they are occupied by a primary user.
%In co-operative scenario however, they might offer an agreement such that a portion of channel frequencies will be accessible.
%\subsection{cognitive radio networks scheduling}
We assume that information about primary user communications are provided reliably.
%(who is active and who is not) we can forget about detection problem in CRNs.
Each primary user has its own channel frequency which is not shared with others.
Thus, primary users can send data over their own licensed
channels to their respective access points simultaneously.
Secondary users do not have such channels and
opportunistically try to send their data to receivers
by utilizing idle primary channels.
Formally we define a CRN by $N$ secondary users and $M$ primary users.
Each secondary user may have access to a subset of primary channels (see Fig.~\ref{fig:cognitive1}) according to their Euclidean distance.
Throughout this paper primary users are denoted by $\{a,b,...\}$ and secondary users by $\{i,j,..\}$.
We define a binary metric ${\cal I}(i,a) =\{0,1\}$ for each secondary user $i\in\{1,...,N\}$
and primary channel $a\in\{1,...,M\}$ where ${\cal I}(i,a) =1$ indicates accessibility of primary channel $a$ to secondary user $i$.

\begin{figure}[htb]
\begin{center}
\hspace{0.4mm}
\includegraphics[width=0.65\columnwidth]{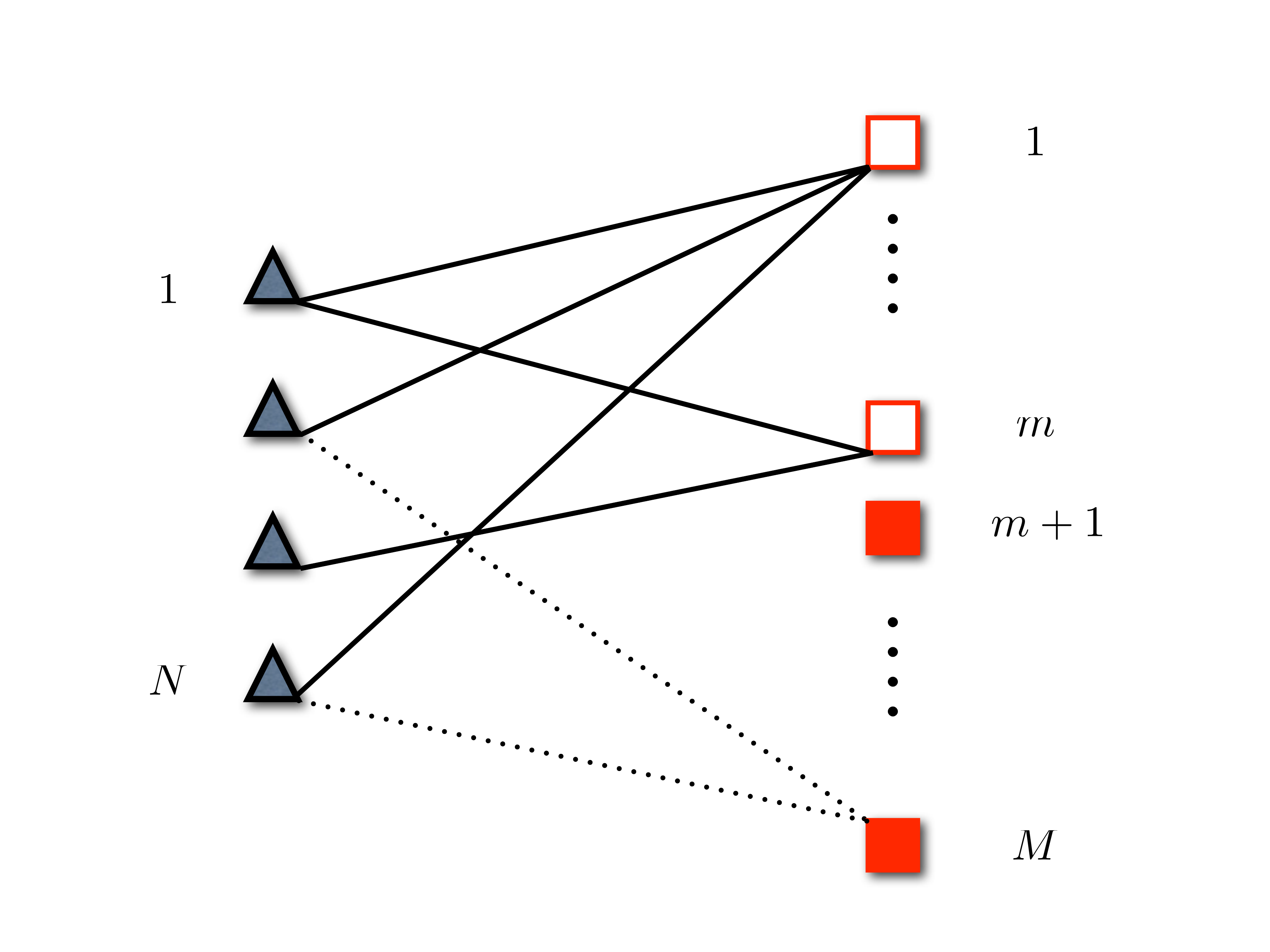}
\vspace*{-1em}
\end{center}
\caption{ A factor graph representation of CR network. Solid links represent the accessibility of the idle primary channels to secondary users (${\cal I}(i,a)$). Dashed lines represent possible interference experienced at active primary users due to transmissions of secondary users (${\cal G}(i,a)$).  }
\label{fig:factor_graph}
\end{figure}
% In~\cite{urgaonkar2008opportunistic} Lyapunov Optimization technique is used to
%design a scheduling policy for both online flow control and resource allocation in CRNs.
%Here,
We present the problem in which message-passing technique can be used to find optimal scheduling in CRNs.
Consider a CRN where a subset of primary users are inactive at each time slot.
Each cognitive user can access a primary channel if the channel is accessible and if its corresponding primary user is inactive.
However, at the same time it should not interfere with transmissions of other active primary users.
We define an interference matrix ${\cal G}(i,a)$
which measures the interfere experienced at primary user $a$ when secondary user $i$ is connected to a free channel.
Note that ${\cal G}(i,a)$ is different from the accessibility matrix ${\cal I}(i,a)$
because a secondary user might not have access to a primary channel (${\cal I}(i,a) =0$) but have a non-zero interference with (active) primary user $a$ (${\cal G}(i,a)\neq 0$) when it is connected to another free channel.
To each secondary user $i$, we assign a binary variable $s_i$ where $s_i=1$ indicates secondary user $i$ connected to
a free channel and $s_i=0$ stands for the case where $i$ is not connected to any primary channel.
We now introduce a factor graph representation of the CRN (Fig.~\ref{fig:factor_graph}). It is a bipartite graph which includes secondary and primary users as different factor nodes, shown by circles and squares, respectively. Secondary and primary users are connected to each other according to the accessibility and interference matrices.
We assume that the first $m$ primary users are idle at the time and only $M-m$ primary users are active (indexed by $(m+1,...,M)$).

In order to distinguish which primary channel is assigned to a secondary user, we
introduce variable $\sigma_{i,a}=\{0,1\}$ for user $i\in \{1,...,N\}$ and its accessible free primary channel $a\in \{1,...,m\}$
 such that $\sigma_{i,a} = 1$ when $i$ is connected to $a$.
By definition we have
\begin{equation}
s_i = \Theta\left(\sum_{a=1}^m \sigma_{i,a}\,\, {\cal I}(i,a)\right)
\label{eq:activity_matrix}
\end{equation}
where $\Theta$ is the step function.
The ultimate goal is to efficiently connect secondary users to free primary channels.
Secondary users might have different priority to be connected according to
their different tasks. The problem then can be expressed as an optimization problem to
 find the best assignment $\vec{s} = \{s_1,...,s_N\}$ such that the following cost function is minimized
\begin{equation}
{\cal H}(\vec{s}\,) = - \sum_{i=1}^N\, s_i c_i
\label{eq:cost}
\end{equation}
Here $c_i$ is the task (priority) of secondary user $i$. Clearly secondary users with higher priority have preference
to be connected. However the interference and accessibility constraints must both be satisfied
which makes the problem hard. Note that it is possible to extend the current prior task vector to
more general cases where secondary users have different priorities to access different unused channels.
%(there we have to work with a matrix instead of vector $c$).

%To identify if a secondary user is connected to an available primary channel we define $\sigma_{i,a}$ as a binary variable between secondary user $i\in \{1,...,N\}$ and its accessible free primary channel $a\in \{1,...,m\}$ with ${\cal I}(i,a) = 1$. A valid communication is then recognized by $\sigma_{i,a} = 1$.
In what follows we describe two different optimization objectives in solving the CRN optimization problem.
The first approach is based on hard constraints induced by active primary users while 
the second is based on soft constraints where
interferences between secondary users and primary users are tolerated
but the cost function is modified accordingly.

%%%%%%%%%%%%%%%%%%%%%%%%%%%%%%%%%%%%%%%%%%%%%%%%%%%
\subsection{Model A}
To formulate the hard interference constraints imposed by active primary users, we introduce a
 quenched variable for every primary user $\vec{\theta} \!= \!\{\theta_1,...,\theta_M\}$
to indicate the maximum interference tolerated. We assume that the sum over interferences from all secondary users connected to free channels must be smaller than the active primary user threshold.
%\begin{equation}
%\sum_{i=1}^N {\cal G}(i,a) s_i \,\, \leq\,\, \theta_a   \,\,\,\,\,\,\,\,\,\,\,\, a\in \{m+1, ... , M\}
%\end{equation}
The optimization problem then is to minimize the cost function Eq.~\ref{eq:cost}
%\begin{equation}
%{\cal H}(\vec{s}\,) = - \sum_{i=1}^N\, s_i c_i
%\end{equation}
%such that following conditions are satisfied :
%%The ultimate task is to find an injective assignment between secondary users and free channels such that
%%For an assignment $\vec{\pi} : \{1,...,i,j,...,N\}\rightarrow \{1,...,a,b,...,M\}$ this is equivalent to minimize the following cost function
%\begin{itemize}
%\item hard interference constrains imposed by active primary users
%\begin{equation}
%  \sum_{i=1}^N  {\cal G}(i,a) s_i \,\, \leq \,\, \theta_a \,\,\,\,\,\,\,\,\,\,\,\, {\rm for}\,\,\, a\in \{m+1,...,M\}
%\end{equation}
%\item each secondary user is connected to at most one free channel at the time
% \begin{equation}
%\sum_{a=1}^m  {\cal I}(i,a)\,\,\sigma_{i, a} \leq 1  \,\,\,\,\,\,\,\,\,\,\,\, {\rm for}\,\,\,i\in \{1,...,N\} \label{eq:constrain1} %{\rm if} \,\,\, \sigma_{a} = 0 \,\,\,\,\,\,\Rightarrow \,\,\,\,\,
%\end{equation}
%\item each free channel is connected to at most one secondary user
%\begin{equation}
%\sum_{i=1}^N  {\cal I}(i,a)\,\,\sigma_{i, a} \leq 1  \,\,\,\,\,\,\,\,\,\,\,\, {\rm for}\,\,\,a\in \{1,...,m\} \label{eq:constrain2}%{\rm if} \,\,\, \sigma_{a} = 0 \,\,\,\,\,\,\Rightarrow \,\,\,\,\,
%\end{equation}
% \end{itemize}
%which must be solved together with Eq.~\ref{eq:activity_matrix}.
such that following conditions are satisfied, which must be solved in conjunction with Eq.~\ref{eq:activity_matrix}:\\
$\bullet~$ hard interference constraints imposed by active primary users
\begin{equation}
  \sum_{i=1}^N  {\cal G}(i,a) s_i \,\, \leq \,\, \theta_a \,\,\,\,\,\,\,\,\,\,\,\, {\rm for}\,\,\, a\in \{m+1,...,M\}
\end{equation}
$\bullet~$ each secondary user is connected to at most one free channel% at the time
 \begin{equation}
\sum_{a=1}^m  {\cal I}(i,a)\,\,\sigma_{i, a} \leq 1  \,\,\,\,\,\,\,\,\,\,\,\, {\rm for}\,\,\,i\in \{1,...,N\} \label{eq:constrain1} %{\rm if} \,\,\, \sigma_{a} = 0 \,\,\,\,\,\,\Rightarrow \,\,\,\,\,
\end{equation}
$\bullet~$ and each free channel to at most one secondary user
\begin{equation}
\sum_{i=1}^N  {\cal I}(i,a)\,\,\sigma_{i, a} \leq 1  \,\,\,\,\,\,\,\,\,\,\,\, {\rm for}\,\,\,a\in \{1,...,m\} \label{eq:constrain2}%{\rm if} \,\,\, \sigma_{a} = 0 \,\,\,\,\,\,\Rightarrow \,\,\,\,\,
\end{equation}

\subsection{Model B}
An alternative approach is to relax the hard interference constraints
between secondary users and active primary channels.
We modify the cost function such that the solution minimizes the interference between secondary users and active primary channels.
We consider a quadratic cost of the interferences between SUs and active PUs.
 \begin{eqnarray}
{\cal H}(\vec{s}\,) = - \sum_{i=1}^N\, s_i c_i + \sum_{a = m+1}^M \, \frac{1}{\theta_a}\left(\sum_{i=1}^N\, {\cal G}(i,a) s_i\right)^2 \nonumber\\
{\cal H}(\vec{s}\,) = - \sum_{i=1}^N\, s_i c_i + \sum_{a = m+1}^M \,\frac{1}{\theta_a} \sum_{i,j=1}^N\, {\cal G}(i,a) {\cal G}(j,a) s_i s_j
\label{eq:modified_cost}
\end{eqnarray}
where the goal is to find $\vec{s}$ that minimizes the cost function (Eq.~\ref{eq:modified_cost}), conditioned on constraints introduced in Eqs.~\ref{eq:constrain1} and~\ref{eq:constrain2}.
This objective represents a trade-off between maximizing the number of connected secondary users and minimizing the induced interference.

In both models, we assume a stationary state described by the Boltzmann distribution
%\begin{equation}
$p(\vec{s}\,) \!=\! \frac{1}{Z} e^{-\beta {\cal H}(\vec{s}\,)} $,
%\end{equation}
where the parameter $\beta$ is analogous to inverse temperature and is a measure of how strictly the minimization in enforced.
In the limit $\beta\to \infty$, it will be concentrated to the minimum of the cost function.
% -------------------------------------------------------------------------------------------------------------------------------------------- BP
\begin{figure*} [htb]
\begin{small}
\begin{eqnarray}
 A_{i\to a}(\sigma_{i,a}) \! \! & \propto & \! \! \sum_{\{\sigma_{i,b}\}\setminus \sigma_{i,a}}\,\, e^{\beta s_i c_i}\displaystyle\prod_{b=1\neq a}^m B_{b\to i}(\sigma_{i,b}) \,\, \displaystyle\prod_{b=m+1}^M D_{b\to i}(s_i) \,\, \,\,\mathds{1}\left(\sigma_{i,a}+\sum_{b=1\neq a}^m \sigma_{i,b}  \,\,{\cal I}(i,b)\leq 1\right) \label{eq:BP_m1}\\
B_{a\to i}(\sigma_{i,a}) & \propto & \sum_{\{\sigma_{j,a}\}\setminus \sigma_{i,a}} \,\, \displaystyle\prod_{j\neq i} A_{j\to a}(\sigma_{j,a}) \,\,\,\, \mathds{1}\left(\sigma_{i,a} + \sum_{j\neq i} \sigma_{j,a}\,\, {\cal I}(j,a) = 1\right) \\
C_{i\to a} (s_i) & \propto &  \sum_{\{\sigma_{i,b}\}}\,\, e^{\beta s_i c_i} \displaystyle\prod_{b=1}^m B_{b\to i}(\sigma_{i,b}) \,\, \displaystyle\prod_{b=m+1\neq a}^M D_{b\to i}(s_i) \,\, \,\,\delta_{s_i,\Theta\left(\sum_{c=1}^m \sigma_{i,c}\,\, {\cal I}(i,c)\right)} \\
%D_{a\to i}(s_i) &\propto& \sum_{s_1,...,s_N} \,\, \displaystyle\prod_{j=1\neq i}^N  C_{j\to a}(s_j) \,\, \,\, \mathds{1}\left(\sum_{j}\, {\cal G}(j,a)\, s_j \leq \theta_a \right)
D_{a\to i}(s_i) &\propto& \sum_{\vec{s}\,\setminus s_i} \,\, \displaystyle\prod_{j=1\neq i}^N  C_{j\to a}(s_j) \,\, \mathds{1}\left(\sum_{j}\, {\cal G}(j,a)\, s_j \leq \theta_a \right)  \,\,\,\,\, {\rm Model \,A}\\
D_{a\to i}(s_i) &\propto& \sum_{\vec{s}\,\setminus s_i} \,\, \displaystyle\prod_{j=1\neq i}^N  C_{j\to a}(s_j) \,\, e^{-\frac{\beta}{\theta_a}   {\cal G}(i,a) s_i \left(\sum_{j=1}^N\, {\cal G}(j,a) s_j\right)} \,\,\,\,\, {\rm Model \,B} \label{eq:BP_m4_B}
\end{eqnarray}
\vspace*{-2em}
\end{small}
\end{figure*}
\section{Belief propagation}
\label{sec:BP}
In a system with a given cost function, the problem of finding the global minimum
or calculating marginal probabilities can be solved approximately by the BP algorithm.
%It was first proposed by Pearl~\cite{pearl1988probabilistic} who formulated
%the BP for loop-less graphs, and was later extended to general graphs. The physical
%interpretation of message-passing algorithms and in particular belief propagation was
%later explained by statistical mechanics studies on disordered systems~\cite{yedidia2001generalized,mezard2009information}.
%
\begin{figure}[htb]
\begin{center}
\hspace{1mm}
\includegraphics[width=0.65\columnwidth]{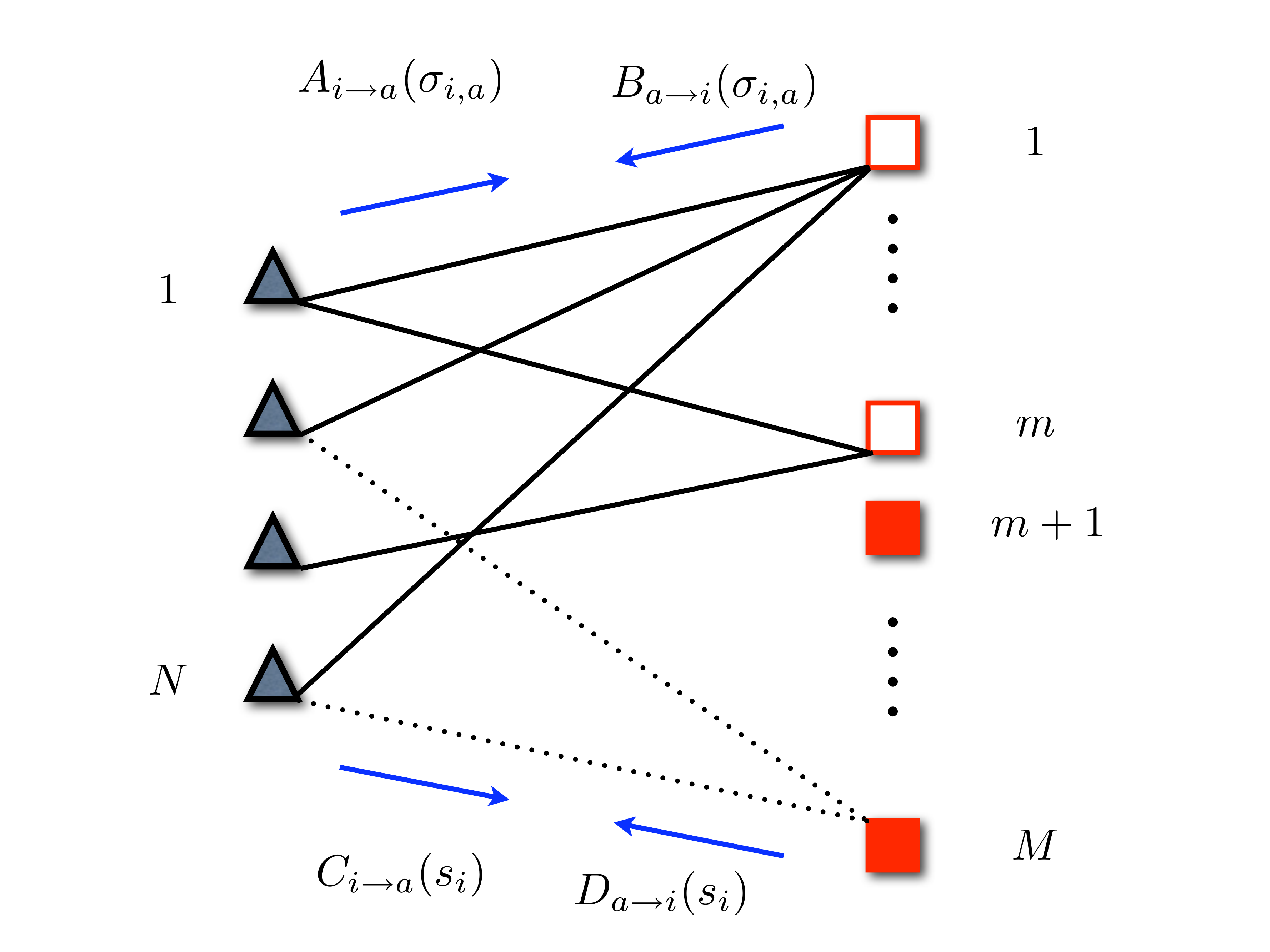}
\vspace*{-1em}
\end{center}
\caption{ Messages are exchanged among secondary users and available primary users and between secondary users and active primary users.
They fulfill a set of consistent equations and correspond to the BP approximation of the corresponding CRN.}
\label{fig:message}
\end{figure}

To implement BP we exchange a set of messages between the different nodes in a factor graph.
The type of messages to be exchanged between secondary users and active primary users are essentially different from those exchanged between secondary users and inactive primary users.
To an active primary user the important information is whether a secondary user with non-zero interference is connected to any free channel, irrespective of the channel.
On the other hand, for free channels it is important to take into account the information about the specific connections (see constraints~\ref{eq:constrain1} and \ref{eq:constrain2}).
We introduce four different messages as follows
\begin{itemize}
\item $A_{i\to a}(\sigma_{i,a})$ is a messages sent from secondary user $i$ to 
free channel $a$ with non-zero accessibility metric ${\cal I}(i,a)\neq 0$.
It gives the probability of observing $\sigma_{i,a}$. % (whether $i$ is connected to $a$ or not).
\item $B_{a\to i}(\sigma_{i,a})$ is the reverse message sent from available primary user $a$ to secondary user $i$.
\item $C_{i\to a}(s_i)$ is a message sent from secondary user $i$ to active primary user $a$ with non-zero interference metric (${\cal G}(i,a)\neq 0$). It gives the probability of secondary user to be connected (not-connected) to one of its available free channel $s_i=1$ ($s_i=0$).
\item $D_{a\to i}(s_i)$ is the corresponding reverse message from active primary user $a$ to secondary user $i$.
\end{itemize}
These BP messages fulfill a set of closed equations (Eq.~\ref{eq:BP_m1}-\ref{eq:BP_m4_B}).

Belief propagation equations are to be solved iteratively. We assign an initial (unbiased) condition to each message
and iterate all messages until a fixed point is reached. This is a fully decentralized procedure and can be applied efficiently.
Once the fixed point is reached, the marginal probabilities for each secondary user can be computed. Each marginal probability represents the probability of connecting corresponding secondary users to a free channel.

Note that in order to update message $D_{a\to i}$ we have to take a summation over $2^{{\cal O}(K)}$ where $K$ is the average connectivity for active primary users.
For sparse interference/connectivity matrices the algorithm provides a fast solution while for dense matrices the  corresponding computational complexity grows
exponentially and becomes infeasible.

For Model B, we will show that computational complexity remains manageable even in the regime where active primary users experience interference from many secondary users. 
To simplify the BP equations we introduce the BP field messages
%\begin{eqnarray}
%h_{i\to a} = \frac{1}{\beta} \,{\rm ln} \left(\frac{A_{i\to a}(1)}{A_{i\to a}(0)}\right) \nonumber \\
%h_{a\to i} = \frac{1}{\beta} \,{\rm ln} \left(\frac{B_{i\to a}(1)}{B_{i\to a}(0)}\right) \nonumber \\
%g_{i\to a} = \frac{1}{\beta} \,{\rm ln} \left(\frac{C_{i\to a}(1)}{C_{i\to a}(0)}\right) \nonumber \\
%g_{a\to i} = \frac{1}{\beta} \,{\rm ln} \left(\frac{D_{i\to a}(1)}{D_{i\to a}(0)}\right)
%\end{eqnarray}
$h_{i\to a} = \frac{1}{\beta} \,{\rm ln} \left(\frac{A_{i\to a}(1)}{A_{i\to a}(0)}\right), h_{a\to i} = \frac{1}{\beta} \,{\rm ln} \left(\frac{B_{i\to a}(1)}{B_{i\to a}(0)}\right), q_{i\to a} = \frac{1}{\beta} \,{\rm ln} \left(\frac{C_{i\to a}(1)}{C_{i\to a}(0)}\right), q_{a\to i} = \frac{1}{\beta} \,{\rm ln} \left(\frac{D_{i\to a}(1)}{D_{i\to a}(0)}\right)$.
%\end{eqnarray}
%
The corresponding new BP equations for model B become
\begin{eqnarray}
h_{i\to a}  \!\!\!\! &=&  \!\!\!\! -\frac{1}{\beta} \,{\rm ln}  \!\! \left[e^{-\beta\left(c_i+\sum_{b=m+1}^M q_{b\to i}\right)}+ \!\! \sum_{b=1\neq a}^m \!\! \!\! e^{\beta h_{b\to i}}\right] \nonumber \\
h_{a\to i}  \!\!\!\! &=&  \!\!\!\! -\frac{1}{\beta} \,{\rm ln} \!\! \left[1+\sum_{j\neq i} e^{\beta h_{j\to a}}\right] \\
q_{i\to a}  \!\!\!\!  &=&  \!\!\!\! - c_i -\! \! \!\!\!\! \!\!\! \sum_{b=m+1\neq a}^M \!\!\!\!  \!\!\! q_{b\to i} -  \! \frac{1}{\beta} \,{\rm ln} \!\! \left[\prod_{b=1}^m \left(1+e^{\beta h_{b\to i}}\right )-1\right] \nonumber \\
q_{a\to i}  \!\!\!\! &=& \!\!\!\! -\frac{1}{\beta}  \! \sum_{j\neq i} \!{\rm ln}\!\! \left[\frac{{\rm exp}(\beta (q_{j\to a}\! - {\cal G}(i,a){\cal G}(j,a)/\theta_a))\!+\!1}{{\rm exp}(\beta q_{j\to a}) +1}\right] \nonumber
\end{eqnarray}
The advantage of this new version of BP is that all equations involve summations over a finite number of terms.
% -------------------------------------------------------------------------------------------------------------------------------------------- results
\section{Simulation results}
\label{sec:simulations}
In this section, we investigate the performance of BP and compare it to a greedy benchmark algorithm. We show how BP outperforms the latter in finding
the better assignment of secondary users to the available free channels.
\begin{figure}[b]
\begin{center}
\hspace{0.4mm}
\includegraphics[width=0.65\columnwidth]{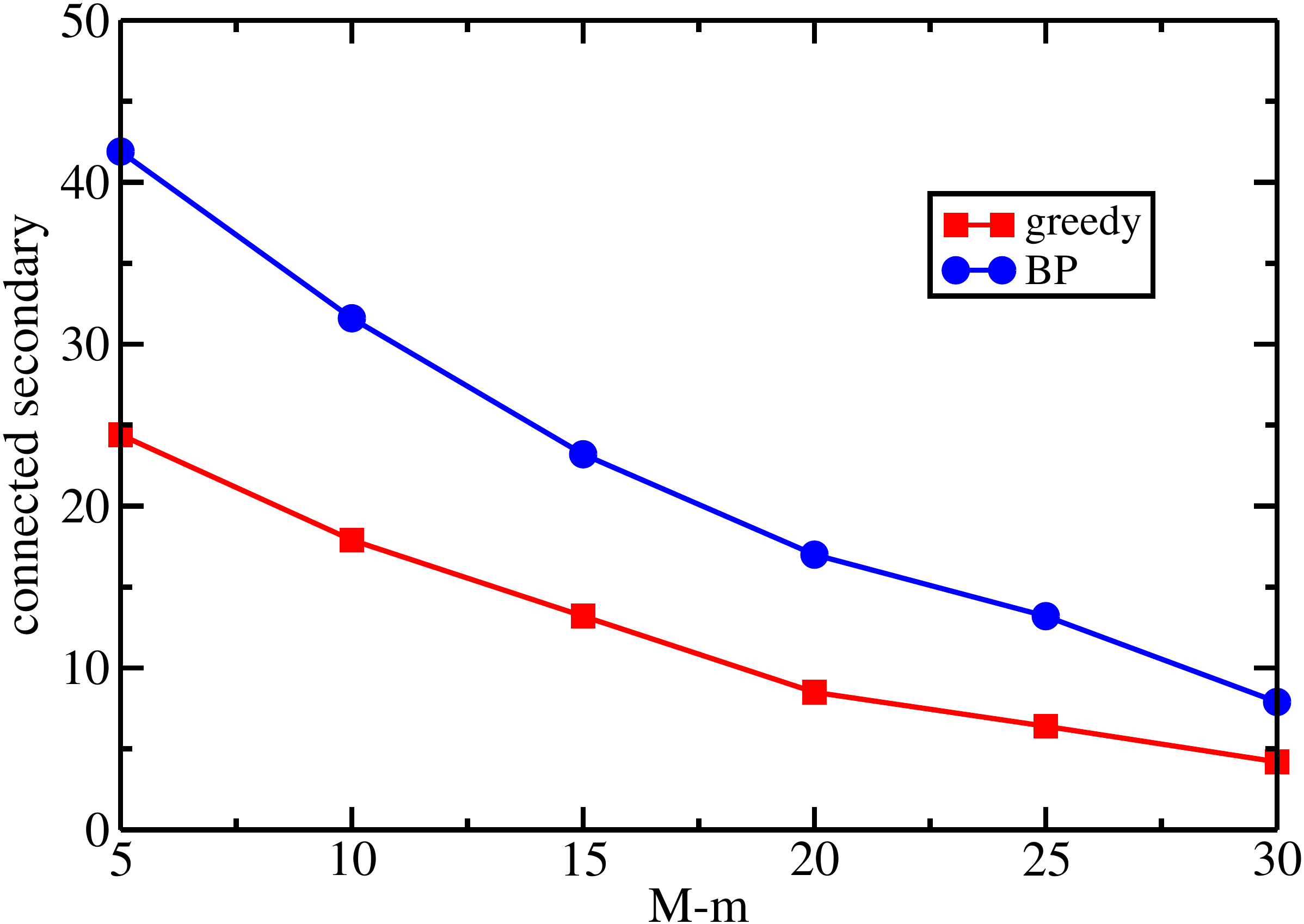}
\vspace*{-1em}
\end{center}
\caption{The number of connected secondary users versus the number of active primary users in a CRN with $N=100$ and $M=50$.
The blue line represents the result obtained by BP while red line represents the performance of the greedy algorithm.
Results are averaged over 10 different network realizations with the same set of parameters. }
\label{fig:n100_mA}
\end{figure}

%------------------------------------------------------------------------------------------------------------------------------------- begin georgios
We consider a cell-based simulation scenario, where base stations are placed uniformly at random
in a unit square (accounting
for irregular urban terrain) and each base station has one associated channel allocated to a primary user.
The secondary users attempt to opportunistically send data to the base-stations using idle primary channels;
we thus concentrate on uplink communication.

Power attenuation at distance $r$ is $g \cdot r^{-\alpha}$, where $\alpha \geq 2$ is the path-loss exponent (we take $\alpha=3.5$)
and $g$ is a non-negative random variable accounting for fading and shadowing, with unit mean. We assume a
Rayleigh fading inspired model, where g is distributed exponentially and is i.i.d. for each secondary
user/base-station pair. We also consider a cut-off value below which the received
power is assumed to be $0$ (this is realistic in practice, and the sum of interference contributions below the threshold can be assumed
to be taken into account as background noise).
Secondary users can connect to a free channel if the SNR (signal to noise ratio) from the
corresponding base-station is above a predefined threshold. Similarly the interference constraint is
a different predetermined threshold. 
%For performance evaluation the defined thresholds are such that the average number of interfering nodes per primary user is approximately $6$.

We also propose a greedy algorithm strategy as a simple benchmark to compare with our message passing
approach. For Model A, the algorithm is the following: we iteratively choose among the remaining links the one
of largest weight (i.e., largest $c_i$) and mark it as active as long as the active links constitute a matching and
the interference constraints are satisfied for all active primary users; otherwise, the link is removed
from the remaining link set.
To solve model B, we apply the same algorithm with a simple modification: for each link we subtract
from its weight the sum of squared interference it generates and omit the interference constraints.
%We thus have a greedy maximal matching algorithm in this case.
%----------------------------------------------------------------------------------------------------------------------------------------end georgios

Through our simulations we set the priority of secondary users to $1$ for all users ($c_i=1$) therefore the absolute value of cost function
in model A is equivalent to the number of connected secondary users (see Eq.~\ref{eq:cost}). The higher this number is the better the algorithm performs. All interference thresholds are also set to $1$ ($\theta_a=1$).
Fig.~\ref{fig:n100_mA} compares results obtained by BP and the greedy algorithm for a cognitive radio networks with $N=100$ secondary and $M=50$ primary users. As we increase the number of active primary users ($M-m$), we end up with less free channel frequency to be shared with secondary users and hence the number of connected secondary users decreases. For all values of active primary users
BP outperforms the greedy algorithm as is shown in the figure.

\begin{figure}[t]
\begin{center}
\hspace{0.4mm}
\includegraphics[width=0.65\columnwidth]{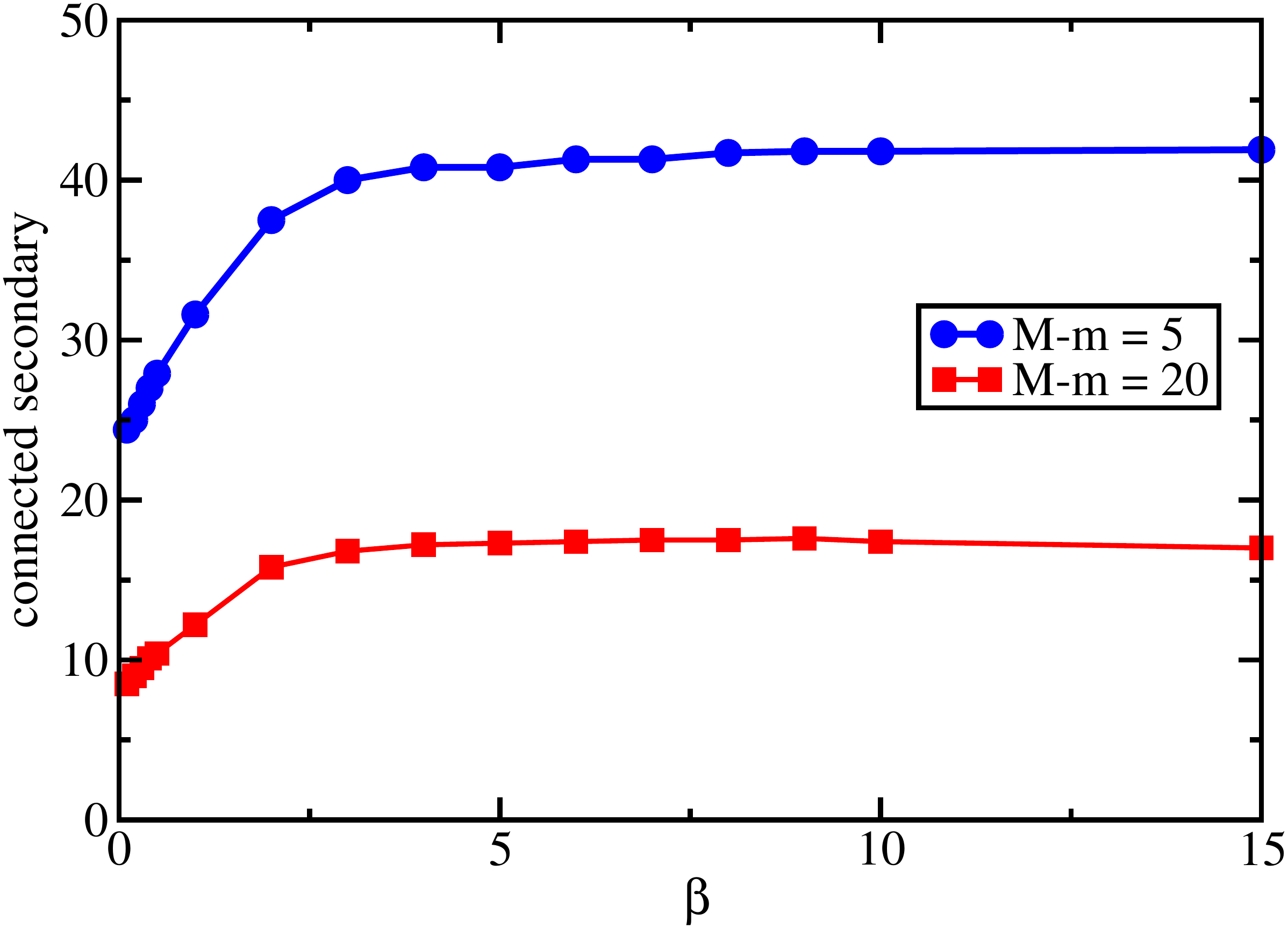}
\vspace*{-1em}
\end{center}
\caption{Performance of BP (Model A) for different values of $\beta$. The results improve by increasing $\beta$
as the optimal solution is expected to be found at $\beta\to \infty$. Simulation results show convergence as $\beta$ increases.}
\label{fig:beta_MA}
\end{figure}

To investigate the optimal solution found by BP we
plot the results obtained for various values of $\beta$ (Fig.~\ref{fig:beta_MA}).
As mentioned earlier, we expect to find the optimal solution by taking the limit $\beta\to \infty$.
However, the number of connected secondary users converges already for $\beta\geq 5$.

\begin{figure}[htb]
\begin{center}
\hspace{0.4mm}
\includegraphics[width=0.65\columnwidth]{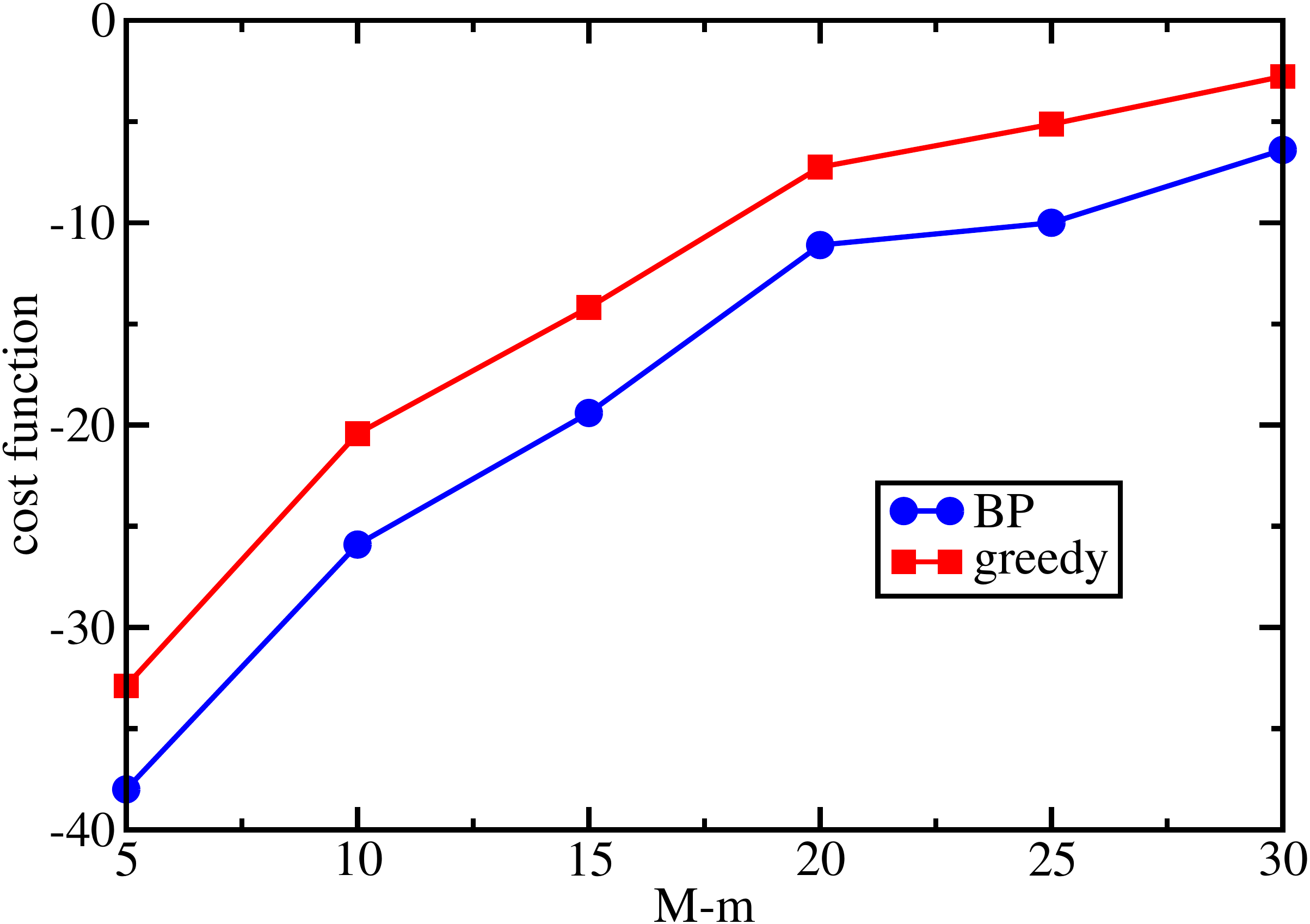}
\vspace*{-1em}
\end{center}
\caption{The minimal cost achieved for model B (Eq.~\ref{eq:modified_cost}) in a cognitive radio networks with $N=100$ secondary and $M=50$ primary users. Results obtained by BP (blue red) show better performance compared to the greedy algorithm (red line). Results are averaged over 10 different network realizations with the same set of parameters.}
\label{fig:n100_mB}
\end{figure}
Similar behavior is observed for Model B with soft interference constraints.
Fig.~\ref{fig:n100_mB} compares results obtained by BP and the greedy algorithm.
It shows the optimal cost value obtained by the different methods for various numbers of active primary users.
For all parameters BP finds a better solution (lower minimal cost) compared to the greedy algorithm.

\begin{figure}[t]
\begin{center}
\hspace{0.4mm}
\includegraphics[width=0.65\columnwidth]{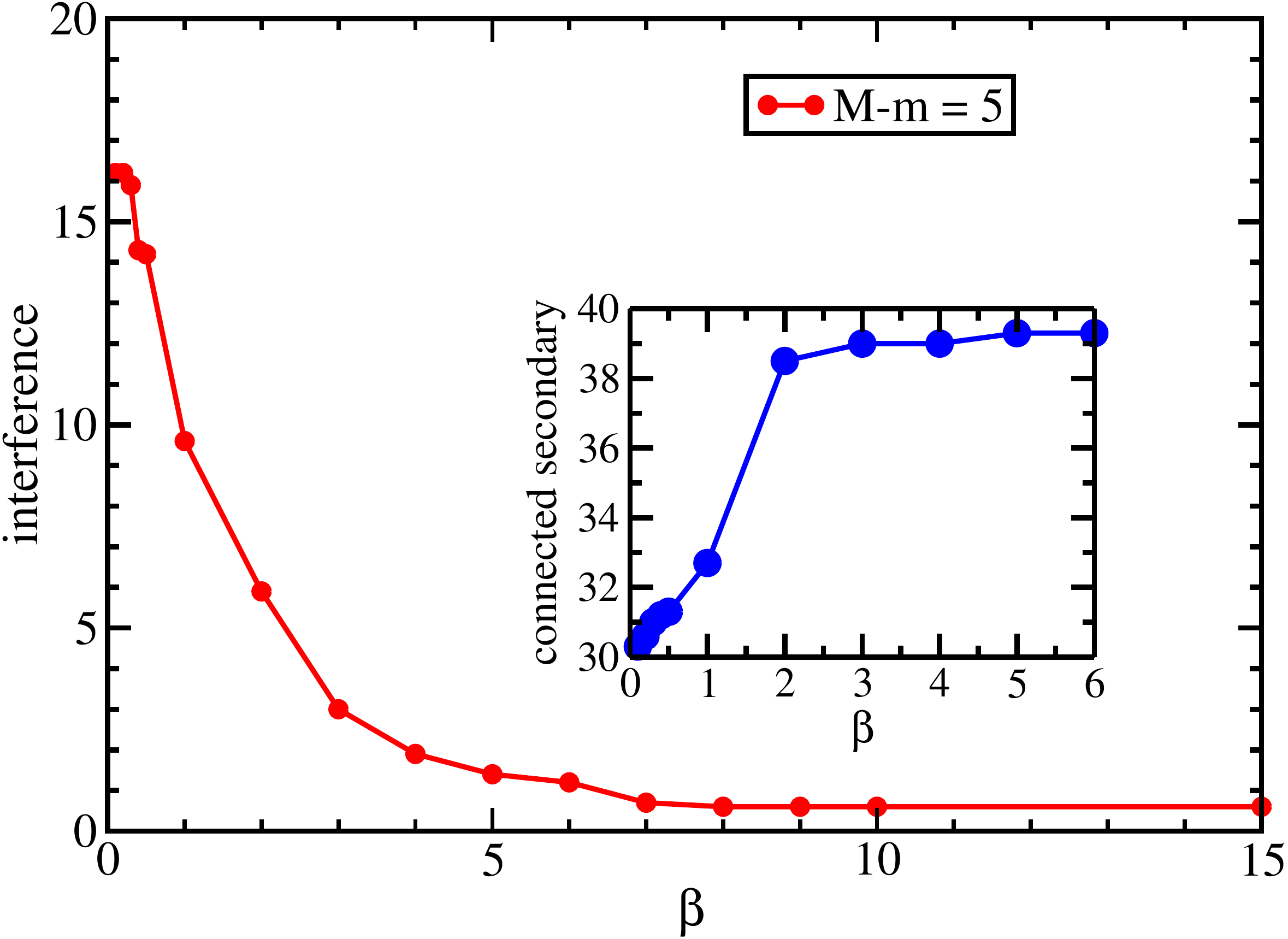}
\vspace*{-1em}
\end{center}
\caption{ The interference between connected secondary and active primary users in a cognitive radio networks with $N=100$, $M=50$ and $M-m$=5. The results are obtained by running BP on 10 different networks realizations. The inset shows the number of connected secondary users. }
\label{fig:beta_MB}
\end{figure}

The performance of BP in model B is examined by monitoring the two parts of cost function: interference and the number of connected secondary users
for different values of $\beta$. As expected, the interference part of cost function ($\sum_{i,j}\sum_{a} G(i,a) G(j,a) s_i s_j$) vanishes
as $\beta$ increases. Figure~\ref{fig:beta_MB} shows the average results for 10 realizations of cognitive networks
with $N=100$, $M=50$ and $M-m=5$. As we increase $\beta$ the interference decreases and the
number of connected secondary users increases.

% -------------------------------------------------------------------------------------------------------------------------------------------- conclusion
\section{Conclusion}
We have applied BP to find the optimal resource utilization in cognitive radio networks.
We have cast the task into a optimization problem where the number of connected secondary users is to be maximized under hard and soft interference constraints. A greedy algorithm is used as a benchmark for the BP results; the 
latter outperforms the former in all the experiments carried out and is computationally efficient.
%---- conclusion georgios:
%We proposed a new Belief Propagation based framework for optimal opportunistic scheduling in cognitive radio networks, taking into account realistic interference constraints.
%We performed simulations to study the improved performance of the BP approach, compared with a benchmark greedy alternative. We have thus initiated a study on the advantages of BP algorithms in CRN scheduling, due to their low complexity, efficiency and fully distributed operation.  Therefore, interesting directions for further research should extend our framework to more complicated scheduling problems, including information theoretic considerations and dynamic network scenarios.

% -------------------------------------------------------------------------------------------------------------------------------------------- refrences
\bibliographystyle{plain}
\bibliography{wireless_ref}
%\begin{thebibliography}{1}
%\end{thebibliography}

\end{document}